# A new non linear mechanism able to generate avalanches based on soil mechanics

## P. Evesque

Lab MSSMat,  UMR 8579 CNRS, Ecole Centrale Paris
92295 CHATENAY-MALABRY, France, e-mail evesque@mssmat.ecp.fr

**Abstract:**
*We propose a general  mechanism based on soil mechanics concepts, such as dilatancy and friction, to explain the fact that avalanches stop at an angle smaller than they start: the mechanism involved is linked to the fact that the stress field near the free surface of a pile built with inclined strata obeys always the plasticity criteria, even when the slope is smaller than the friction angle. It results from this that the larger the slope angle the smaller the mean stress and the smaller the maximum principle stress. So when the pile rotates to generate the next instability the granular material is submitted to a decrease of the mean stress so that its apparent friction angle $\varphi_{peak}$ increases, becoming larger than the friction angle $\varphi$ . The slope starts then flowing at an angle larger than the friction angle.*

______________________________________________________________________________

Nowadays, a great number of works deal with the problem of sand avalanches [1-10]. Part of them are investigating the flow characteristics [3, 4], others try to understand the limit of stability of an inclined slope [6-9], others try to study 1/f noise generation [5, 6, 7]; but an intriguing question remains: why does the surface flows intermittently and does not flow continuously when the pile is rotated slowly? This seems a trivial question at first sight, and one can give a simple explanation: the friction coefficient depends of the flow speed, and it present a minimum for a small but finite speed [2]; it results from this hypothesis that the static friction coefficient is not the smallest one and that it exists a maximum angle of repose which is larger than the dynamics one.

However this explanation cannot satisfy completely the specialist of granular matter for whom the quasi-static regime does exist, because it has been recognised from long using triaxial apparatus or shear cells that friction coefficient does not depend on the speed of the deformation as far as it remains small enough. This is also in agreement with recent results on shear experiments [10] by physicists. So there shall be an other explanation.

Application of the principles of the mechanics of granular material allows to understand the existence of avalanches as generated by a dilatancy mechanism [6, 7]. Indeed, this mechanism modifies the apparent fiction angle and makes it larger than the real one, i.e. the "critical" one; as in [7], we use the term "critical" to make reference to the critical state of soil mechanics as defined in Schofield & Wroth [11] for the critical state and to distinguish it from a true critical behaviour (in the sense of phase transition). Extended investigation [8-9] of the properties of avalanches using centrifuge confirmed the importance of soil mechanics concepts and have demonstrated the relevance of the pile density and of the gravity as two main parameters controlling the size of the avalanches. However, simple analysis using soil mechanics concept predicts that it should be possible to get the size of the avalanche to tend to zero systematically after the generation of few macroscopic avalanches in a rotating container from which avalanches flow outside [6,7,9]. Despite this prediction, it was not possible to get a continuous-flow regime with a flow tending to zero, except when running the experiment in some very peculiar conditions and during a short lapse of time [9]: a continuous avalanche flow could be got in such a way that one could stop it and restart it at will just by stopping the rotation of the container and restarting it; but this requires very specific conditions [9].

So soil mechanics specialists are then faced to a dilemma.

This paper is aimed at raising it and at trying presenting a new explanation of the existence of avalanche process which is based on pure soil mechanics approach. The idea underlying this new understanding was caused by the study of the stress distribution in a conic or a triangular pile using computer simulation [10] with elasto-plastic modelling. This study shew that the plastic criterion was always satisfied near the free surface when piles were generated by layering of inclined strata whose surface orientation is parallel to the free surface $\alpha$. This result holds true whatever the pile-slope inclination is i.e. just at the friction angle $\alpha=\varphi$ or much below $\alpha<\varphi$. However, this results does not hold true when building the pile using horizontal strata, since in this case the pile surface becomes only unstable just for $\alpha=\varphi$.

The idea to understand the mechanism of avalanche uses this result. Let us consider a pile with a slope angle $\alpha$ smaller than $\varphi$ but built according to inclined-strata process. And let us consider an inclined slice of material at the free surface and parallel to it; its thickness is $\delta$ and its surface is S. Let $\sigma_1$ S and $\tau_1$ S be the components of the force exerted by this slice of material on the material located below it; so equilibrium implies :

$$\tau_1 = \rho\ g\ \delta\ \sin(\alpha) \quad (1.a)$$

$$\sigma_1 = \rho\ g\ \delta\ \cos(\alpha) \quad (1.b)$$

As the real stress field corresponds to the plastic condition the stress tensor can be characterised by a Mohr circle which is tangent to the line $\tau = \sigma \tan(\varphi)$ as sketched in Fig. 1; any circle of this kind is characterised by a radius $(\sigma_1-\sigma_2)/2$ and a centre $(\sigma_1+\sigma_2)/2$ located on the $\sigma$ axis such as:

$$(\sigma_1-\sigma_2)/(\sigma_1+\sigma_2)=\sin(\varphi) \quad (2)$$

so that the equation of the circle is of the kind:

$$\tau^2 + (\sigma-(\sigma_1+\sigma_2)/2)^2 = (\sigma_1-\sigma_2)^2/4 = (\sigma_1+\sigma_2)^2 \sin^2(\varphi)/4 \quad (3)$$

**States wich correspond to plastic limit:**

***1- Centre of the Mohr circle and mean stress:***

The adequate Mohr circles, which characterises the stress field of the slope $\alpha$

verifying the plasticity criterion correspond to the two circles which pass through the point of coordinates given by Eq. (1) and which is tangent to the line $\tau = \sigma \tan(\varphi)$. So, replacing $\tau$ and $\sigma$ by their values of Eq. (1) into Eq. (3) allows to find the abscissa of the Mohr circle centre $\sigma_\alpha = (\sigma_1+\sigma_2)/2$:

$$\sigma_\varphi = (\rho\ g\ \delta)/\cos(\varphi) \qquad \text{when } \alpha=\varphi \qquad (4.a)$$

$$\sigma_\alpha = (\rho\ g\ \delta)[\cos(\alpha)\pm\{\cos^2(\alpha)-\cos^2(\varphi)\}^{1/2}]/\cos^2(\varphi) \qquad (4.b)$$

$\sigma_\alpha$ is also the mean stress supported by the plane inclined at $\alpha$. So, there are two such solutions for this circle; the first one (the second one) corresponds to a circle whose centre corresponds to a mean stress $\sigma_\alpha$ larger (smaller) than the mean stress $\sigma_\varphi$ the system would sustain if it was just at the limit of equilibrium when $\alpha=\varphi$.

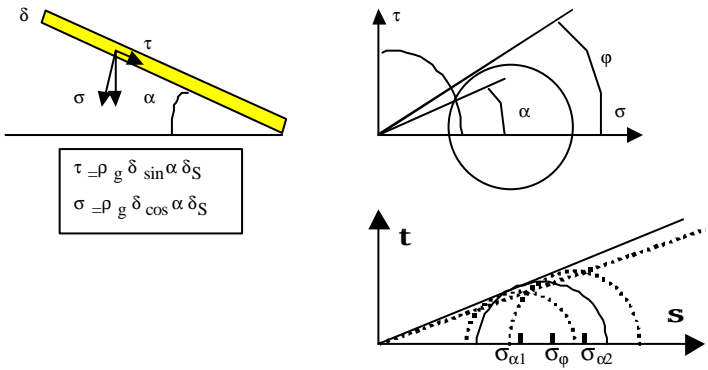

**Figure 1:** *sketch of the slope free surface; and sketch of one of the two Mohr-Coulomb circles which correspond to plastic criterion ; in this plane, any stress tensor is represented by a Mohr-Coulomb circle whose centre is on the $\sigma$ axis. So, the Mohr-Coulomb circles which can represent the stress field near the surface passes through the point defined by $(\tau_1,\sigma_1)$ and cannot cross the line $\tau=\tan(\varphi)\ \sigma$.*

Developing Eq. (4.b) to leading order as a function of $(\alpha-\varphi)$, in the limit of $(\alpha-\varphi)$ tending to zero, one gets the mean stress to be:

$$\sigma_\alpha \cong \sigma_\varphi [1 \pm \{2\ (\varphi-\alpha)\ \tan(\varphi)\}^{1/2} + (\alpha-\varphi)\tan(\varphi)] \qquad (5)$$

This Eq. (5) shows that $\sigma_\alpha$ varies very quickly with $\varphi-\alpha$ in the vicinity of $\varphi$ since it scales as $(\varphi-\alpha)^{1/2}$. This means also that the variations of $\sigma_\alpha$ obeys the law followed by critical bifurcation at $\alpha=\varphi$. This is in agreement with what we have discussed previously in [6,7].

## 2- Principal directions

Principal directions of stress can be obtained by determining the angle between the $\sigma$ axis of the Mohr Coulomb plane and the vector which starts from the centre of the Mohr circle and the point which characterises the force on plane inclined at α, i.e. $(\sigma_1, \tau_1)$ : half of this angle is equal to the angle between the normal to the free surface and the main principal direction of stress.

Labelling u this angle, one shall have, using Eq. (2):

$$\tau_1 = \sin(\pi-u)\,(\sigma_1-\sigma_2)/2 = \sin(\pi-u)\,\sin(\varphi)\,\sigma_\alpha \qquad (6)$$

As $\tau_1$ is given by Eq. (1.a) and $\sigma_\alpha$ by Eq. (4.b), this leads to:

$$\sin(\pi-u) = \sin(\alpha)\cos^2(\varphi) / [\sin(\varphi)\{\cos(\alpha) \pm [\cos^2(\alpha)-\cos^2(\varphi)]^{1/2}\}] \qquad (7)$$

Furthermore, when α=φ, one has u=π/2+φ. So, one can perform a development limited around this value. Labelling $\pi-u = \pi/2+\varphi+\delta\omega$ or $u = \pi/2 +\varphi -\delta\omega$ one gets at leading order:

$$\delta\omega \cong \pm\{2(\varphi-\alpha)\cos(\varphi)/\sin(\varphi)\}^{1/2} \qquad (8)$$

The sign in front of Eq. (8) is the sign in Eq. (7) and in Eq. (4b); it is positive (negative) if $\sigma_\alpha$ is larger (resp. smaller) than $\sigma_\varphi$. As the angle between the normal to the free surface and the principal direction of stress is u/2, the angle β between the vertical direction and the principal direction of stress is $\beta = u/2-\alpha = \pi/4+\varphi/2+\delta\omega/2-\alpha$. This means at leading order:

$$\beta = u/2-\alpha \cong (\pi/4-\varphi/2) \pm \{(\varphi-\alpha)/[2\tan(\varphi)]\}^{1/2} \qquad (9)$$

the second term being positive (negative) when $\sigma_\alpha$ is larger (*resp.* smaller) than $\sigma_\varphi$, so that main principal direction is more (*resp*. less) inclined compared to vertical than the stability limit value, i.e. $\beta_s = \pi/4-\varphi/2$, and so that the sliding direction is less (*resp.* more) inclined compared to horizontal than the sliding angle φ.

### 3- *Dilatancy effect:*

Eq. (5) indicates that $\delta\sigma/\sigma$ varies fast around α=φ ; so the mean stress for α<φ can be either much larger or much smaller than the one for α=φ, since $\delta\sigma/\sigma$ scales as $(\alpha-\varphi)^{1/2}$.

Hence, let us assume for a while that the slope was built at α<φ, and let us also assume that the stress $\sigma_\alpha$ the material of the slope is submitted to is the larger one; it is then larger than $\sigma_\varphi$; the material density $d_{c\alpha}$ which corresponds to this stress is then larger than $d_{c\varphi}$ which is the density of the state just at limit of equilibrium when α=φ. In this case, one expects a dilatancy effect to occur when rotating the slope to get α=φ since the pile has been densified at $d_{c\alpha} > d_{c\varphi}$ during the previous period when it was submitted to the large value of $\sigma_\alpha$ ; so, this dilatancy generates avalanche regime spontaneously.

Let us now consider the case when the mean stress $\sigma_\alpha$ of the material is the smaller one. In this case, one gets the contrary, i.e. $d_{c\alpha} < d_{c\varphi}$. Rotating the slope makes the stress to increase and $d_{c\alpha}$ to increase too. This generates a flow which shall get more and more unstable due to the increase of $d_{c\alpha}$ for a tiny change of angle α. We expect then that the flow can stop at any time when stopping rotating the drum under these circumstances, but we expect also that the apparent friction angle increases faster than the slope angle $\alpha$. This last point leads to predict that flow occurs with an amplification factor.

Both effects were already reported in literature [9], the first one being the most common, but the second one which implies amplification and rotation induced flow were also reported in [9].

## 4- Further advance in stability analysis using classical rules of soil mechanics

Let us describe further the non linear behaviour. A simple rule of soil mechanics relates the void index e to the apparent friction angle $\varphi_{peak}$ [12]:

$$e \tan(\varphi_{peak}) = e_c \tan(\varphi_c) \tag{10}$$

It is recalled that the specific volume v is related to the void index e by $v=1+e$; here $e_c$ is the void index of the critical point. It is known also [12, 13] that $e_c$ depends on the stress field $p=(\sigma_1+2\sigma_3)/3$ :

$$e_c = e_{co} - \lambda \ln(p/p_o) = e_{co} - \lambda \ln(\sigma_3/\sigma_{3o}) \tag{11}$$

Eq. (11) imposes that this Eq. holds true also for $\sigma_\alpha$ and $\sigma_\varphi$. So one deduces from Eq. (11) that $\delta e = e_{co} - e_c$ varies as:

$$\delta e_c = \lambda \ln(p/p_o) = \lambda \ln(\sigma_\alpha/\sigma_\varphi) \tag{12}$$

Further more taking the limit of small difference between $\sigma_\alpha$ and $\sigma_\varphi$, and using Eq. (5) in Eq. (12), one gets:

$$\delta e_c = \pm \lambda \{2 (\varphi-\alpha) \tan(\varphi)\}^{1/2} \tag{13}$$

In turn, $\delta e_c$ generates an apparent friction angle $\varphi_{peak}$ which is different from the friction angle $\varphi$, since from Eq. (10), one gets

$$(\varphi_{peak}-\varphi) = \delta e [\sin(\varphi) \cos(\varphi)]/e_c \tag{14}$$

$$(\varphi_{peak}-\varphi) = \pm \lambda \sqrt{2} (\varphi-\alpha)^{1/2} [\sin(\varphi)^{3/2} \cos^{-1/2}(\varphi)]/e_c \tag{15}$$

In principle Eq. (10) is satisfied experimentally for $\varphi_{peak}$ larger than $\varphi$. However, let us assume that it can be extrapolated to value of $\varphi_{peak}$ smaller than $\varphi$. In such a case, Eq. (15) leads to stable solution if $\varphi_{peak} > \alpha$ only, otherwise the real angle of the slope, i.e. $\alpha$, becomes larger than $\varphi_{peak}$. So, possible solutions are only those ones which correspond to a positive sign in Eq. (15). This means that the mean stress $\sigma_\alpha$ shall be larger than the critical one. This explains why avalanches are generated in a rotating box experiments.

However, this explanation, which assumes the validity of Eq. (10) for $e<e_c$ and $e>e_c$, means that the hypothesis which we used in the last paragraph of section 3 is not correct since the whole slope should be unstable when $\sigma_\alpha < \sigma_\varphi$.

So, one should invoke in this case some non uniformity of the slope in order to explain the results of flow amplification and of stoppage of the flow by stopping rotating the slope. This is a possibility since undulation of the slope perpendicular to the direction of the steepest descent was observed experimentally [9] .

**Conclusion:**

It is proposed in this paper a pure soil mechanics mechanism to explain why the regime of avalanches is spontaneously generated in rotating box experiments. The idea underlying the proposed mechanism has been initiated by a computer result, which has demonstrated that the layer near the free surface is at limit of plasticity. This modelling invokes also the dilatancy mechanism. It shows that this dilatancy mechanism is spontaneously generated since if the flow stops for a slope slightly smaller than the critical angle this is enough to provoke a spontaneous densification process of the pile surface which generates in turn a dilatancy mechanism. This forces the pile to become unstable for an apparent friction angle $\varphi_{peak}$ larger than $\varphi$. When the slope instability is reached, i.e. $\alpha = \varphi_{peak}$, the avalanche proceeds; the flow stops when $\alpha \cong \varphi$; but $\alpha \leq \varphi$ when the material flows outside the container [13]. In general $\alpha$ is slightly smaller than $\varphi$; however due to the non linear character of the densification process as demonstrated by Eq. (15), i.e. $\varphi_{peak} - \varphi \approx \sqrt{(\varphi-\alpha)}$, this produces a macroscopic effect even when $\alpha \cong \varphi$. This generates in turn the new avalanche.

It is expected that the faster the flow the larger the dilatancy effect, i.e. the larger $\varphi_{peak} - \varphi \approx \sqrt{(\varphi-\alpha)}$, so that one expects that the avalanche size be larger in air than in water. This was observed indeed very recently [10].

*Acknowledgements:* CNES is thanked for partial funding.

The electronic arXiv.org version of this paper has been settled during a stay at the Kavli Institute of Theoretical Physics of the University of California at Santa Barbara (KITP-UCSB), in june 2005, supported in part by the National Science Fundation under Grant n° PHY99-07949.*Poudres & Grains* can be found at :
http://www.mssmat.ecp.fr/rubrique.php3?id_rubrique=402

*poudres & grains*